\documentclass[showpacs,prb,twocolumn]{revtex4}%,draft
\usepackage{amsmath}
\usepackage{amssymb}
\usepackage{color}
\usepackage{graphics}
\usepackage{epsfig}
\usepackage{graphicx}
\def\boldrm#1{{\bf #1}}

\def\q{\quad}

\def\p{ \partial }

\begin{document}

\title {Scattering of Surface Plasmon Polaritons by one-dimensional inhomogeneities.}

\author{ A.~Yu.~Nikitin$^{1,2}$}
\email{alexeynik@rambler.ru}
\author{F.~L\'{o}pez-Tejeira$^1$}
\author{ L.~Mart\'{\i}n-Moreno$^1$}
\email{lmm@unizar.es}
 \affiliation{$^1$ Departamento de F\'{\i}sica de la Materia Condensada-ICMA,
Universidad de Zaragoza, E-50009 Zaragoza, Spain \\
$^2$ Theoretical Physics Department, A.Ya. Usikov Institute for
Radiophysics and Electronics, Ukrainian Academy of Sciences, 12
Acad. Proskura Str., 61085 Kharkov, Ukraine}

\begin{abstract}
The scattering of surface plasmons polaritons by a one-dimensional defect of the surface is theoretically studied, by means of both Rayleigh and
modal expansions. The considered defects are either relief perturbations or variations in the permittivity of the metal. The dependence of
transmission, reflection and out-of-plane scattering on parameters defining the defect
 is presented.
We find that the radiated energy is forwardly directed (with
respect to the surface plasmon propagation) in the case of an
impedance defect. However, for relief defects, the radiated energy
may be directed into backward or forward (or both) directions,
depending on the defect width.
\end{abstract}

\maketitle

\section{\label{sec:Intro}Introduction}

The study of electromagnetic excitations localized on
metal-dielectric interfaces, or surface plasmon polaritons (SPPs),
has become of essential importance due to its potentiality for the
implementation of sub-wavelength photonic
circuits.\cite{Ebbesen_Nature_03,Plasmon_devices_05} As a result of
such study, phenomena like total suppression of
reflection,\cite{TSSR_Hutley76} transformation of
polarization,\cite{TP_Hutley90,KatsPRB,KNN_PhysRev} enhanced light
transmission,\cite{Ebbesen98_Nature,Ebbesen_theory_PRL01} formation
of
band-gaps\cite{FullGapBarnes_PRL96,Bozhe_PRL,SPIE_2D03,Maradud_2D_gaps_PRB02}
and other interesting properties of plasmonic crystals have been
discovered. Lately, a great deal of attention has been devoted to
the creation of optical elements for
SPPs,\cite{Ebbesen_45mirrors_06,Dereux_SPwaveguides_PRB04,Krenn_micro-nano-03,
Krenn_metal_stripes-PRB01,EbbesBozhev_inter_nat06}
as well as to the efficient coupling of light into and out of SPPs.
These latter problems require a precise knowledge of the scattering
coefficients of the dispersion centers (i.e. deviations from a flat
metal-dielectric interface) placed on the path of SPPs. However,
while many theoretical works have studied the problem of SPP
scattering by rough surfaces, scattering from simple geometries is
not so well known. This was undoubtedly due to the previous lack of
reproducible experimental data, which are now available thanks to
recent advances in the controlled patterning of metal surfaces.

\begin{figure}[]
\begin{center}
\includegraphics[]{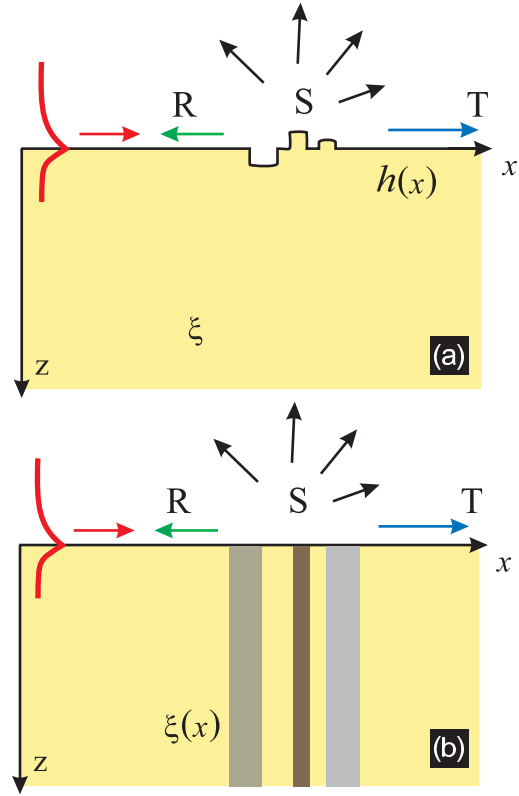}
\end{center}
\caption{\label{geometry}(Color online) Schematic illustration of
the studied system: SPP scattering at the inhomogeneity formed by
either (a) the perturbation of the interface profile $h(x)$ or (b)
the surface impedance of the metal $\xi(x)$.}
\end{figure}

From the theoretical side, the calculation of electromagnetic (EM)
fields on a metal surface in the optical regime is a well-defined
but difficult problem. Although the macroscopic Maxwell equations
govern very accurately the interaction of the EM fields with the
solid, their solution is difficult due to the different ranges of
length scales involved (system size, wavelength, skin depth etc.).
Several techniques have been applied to this problem, each of them
with their advantages and drawbacks. The Greens's Dyadic
technique\cite{MaradudGreffet_PRB94,Olivier,BozhDyadic_PRB} or the
Discrete Dipole
Approximation\cite{Draine,Andrei_APB06,Andrei_LPL06} are virtually
exact methods which suffer from the large (quite often
prohibitive) numerical cost involved with inversion of huge
matrices and the need to calculate cumbersome Sommerfeld
integrals. On the other hand, the mode matching technique is
computationally much simpler, but it can only be directly applied
to the case of indentations in a metal film (and not to
protrusions).\cite{Luis_mirrors_05,Luis_modes_PRL04} Additionally,
it requires the use of surface impedance boundary conditions
(SIBC), which are only applicable when the metallic dielectric
constant, $\epsilon$, satisfies $|\epsilon| \gg 1$. An alternative
approach based on the Rayleigh approximation (which is valid for
small scatterers) has been also extensively applied in the context
of EM scattering by rough surfaces.\cite{Nano-optics_PR05} Even
within this approximation, the calculation of the scattering
coefficients requires solving a difficult integral
equation.\cite{RoughSurfacePRB77,Tatarskii_JOSA,Shchegrov_PRL97}
Several works have rendered this integral equation into a more
manageable form by means of an additional approximation, which
assigns a geometry-dependent ``local impedance'' to the surface
relief (see, for instance, Refs. \onlinecite{MaradudinIBC,
SanchesAppl.Lett.98,MaradudinPRB99}). Up to our knowledge, all
works have concentrated on scatterers with translational symmetry
in one direction, onto which the SPP impinges at normal incidence,
except for the case of a circularly symmetric defect, considered
in Ref.~\onlinecite{Shchegrov_PRL97} by using the reduced Rayleigh
equation.

In this work, we use the Rayleigh approximation combined with the
SIBC at the metal-vacuum interface, systematically taking into
account the geometry of the surface. Our first result is an
integral equation governing the SPP scattering by arbitrary
perturbations, which can be either spatial variations in metal
permittivity or in surface profile. This integral equation is as
simple as the one defined by a ``local
impedance'',\cite{MaradudinIBC,SanchesAppl.Lett.98,MaradudinPRB99}
but is far more accurate, as shown by comparisons with results
obtained using the mode-matching technique. Due to its simplicity,
the method developed provides a clear physical description of SPP
scattering effects. As a first application, here we concentrate on
the scattering properties of a single inhomogeneity, as a function
of shape and geometrical parameters defining the defect.

\section{\label{sec:setup}Rayleigh expansion approximation}
\subsection{Fields representation and boundary conditions}

Consider a monochromatic surface plasmon with frequency $\omega$,
impinging along a metal-vacuum interface at normal incidence (with
in-plane wavevector $k_p$) onto an inhomogeneity region, see
Fig.~\ref{geometry}. In keeping with other works on scattering of
SPPs, we consider that the metal is lossless. From a physicist's
point of view, we expect this to be valid in the systems we have in
mind, as far as the dimension of the inhomogeneity in the direction
of SPP propagation is smaller than the absorption length. More
mathematically, SPP scattering channels are well defined in a
lossless metal, and current conservation provides a strong (although
not definitive!) test for the theory. Nevertheless the proposed
method could, mutatis mutandis, accommodate for absorption. In this
case, instead of scattering coefficients, the outcome of the
calculation would be spacial EM field distributions which could be
used, for instance, to analyze scanning near-field optical
microscope or leakage radiation experiments.

The considered inhomogeneities can be due to variations either in
the surface impedance or in the surface profile. Although the
integral equations governing scattering will have the same
functional form for both cases, the derivations are slightly
different. Let us concentrate first in the latter case, which is
somewhat more involved. Suppose that the metal has a dielectric
constant $\epsilon$ (and correspondingly a surface impedance $\xi
= 1/\sqrt{\epsilon}$) and that the surface relief profile has the
functional form
\begin{equation}\label{0}
z = h(x), \q h(x) = \int dk\, h(k)\exp(ikx).
\end{equation}
Assuming that both the variation of the surface relief and its derivative
are small ($|h|\ll\lambda$ and $|\p_{x}h|\ll1$), we may represent the field
over the surface in the form of Rayleigh expansion.\cite{Book_waves} The
non-zero EM field components in the vacuum half-space can be expanded in
terms of the incident SPP plus scattered field as
 %\begin{equation}\label{2}
\begin{eqnarray}\label{2}
 \left\{
\begin{array}{c}
   E_x(x,z) \\
   E_z(x,z) \\
   H_y(x,z) \\
 \end{array}
 \right\}  &=&
  \left\{
\begin{array}{c}
   -k_{zp}/g \\
   -k_{p}/g \\
   1 \\
 \end{array}
 \right\}  \exp\left[i(k_{p}x - k_{zp}z)\right] \\ \nonumber
 & &
 + \int dk
 \left\{
 \begin{array}{c}
   E_{x}(k) \\
   E_{z}(k) \\
   H_y(k) \\
 \end{array}
 \right\}\exp\left[i(kx-k_{z}z)\right] .
\end{eqnarray}
%\end{equation}
where $ g=\omega/c =2\pi/\lambda$ , $ k_{z}=\sqrt{g^2-k^2}$ and
 $ k_{zp}=\sqrt{g^2-k_{p}^2} =
-\xi g$. The branch of the square root should be chosen such that $ \mathrm{Im}(k_{z})\geq0$, in order to satisfy the
radiation condition. Notice that, within the SIBC, the SPP dispersion relation is $k_p= g(1-\xi^2)^{1/2} $, which
approximates the exact SPP dispersion relation $k_p= g[\epsilon/(1+\epsilon)]^{1/2} $ at $|\epsilon|\gg 1$.

Within the SIBC,\cite{Landau} the EM fields should satisfy
\begin{equation}\label{4}
\mathbf{E}_{t}(x,z) = \xi   \mathbf{H}_{t}(x,z)\times\boldrm{n}(x) ,
\, \mathrm{at} \, z=h(x) ,
\end{equation}
 where $\boldrm{n} = (n_x,
0, n_z)$ is the unitary vector normal to the surface (directed
into the metal half-space) and subscript $t$ corresponds to the
tangential components of the fields.

Notice that the SIBC assumes that the radius of curvature of the
surface is much larger than the skin depth. However, as the
comparison with results obtained by the modal expansion method
will show, the SIBC still represents accurately scattering by
defects where this condition is not fulfilled at a small number of
points (as occurs for rectangular and triangular indentations).

\subsection{The integral equation for the scattered field amplitudes}

Expressing the tangential component of the electric field as
\begin{equation}\label{r1}
\mathbf{E}_{t}(x,z) = \mathbf{E}(x,z) - \boldrm{n}(x)[\mathbf{E}(x,z)\cdot \boldrm{n}(x)],
\end{equation}
the $x$ component of Eq. \eqref{4} gives
\begin{equation}\label{r2}
E_{x}(x,z)n_z(x)-E_{z}(x,z)n_x(x) = \xi H_y(x,z), \, \mathrm{at} \, z=h(x)
\end{equation}
Substituting the fields from Eq.~\eqref{2} into Eqs. \eqref{r2}, and using Maxwell equations $E_x=-(\imath/g) (\partial/\partial z)H_y$ and
$E_z=(\imath/g)(\partial/\partial x)H_y$, provides an integral equation for the amplitudes of the scattered fields, for instance, for the magnetic
component, $ H_y(k)$. A much more manageable equation can be obtained for smooth surface inhomogeneities by expanding the boundary condition over the
small parametes $|\p_{x}h|,|h|/\lambda\ll1$. As the surface normal vector has the form
\begin{equation}\label{r4}
\mathbf{n}=\frac{\mathbf{e}_z-\mathbf{e}_x\p_{x}h}{\sqrt{1+(\p_{x}h)^2}},
\end{equation}
the expansion of these vector components up to second-order terms gives
\begin{eqnarray}\label{r5}
%\begin{equation}\label{r5}
%\begin{split}
n_z &=& 1-\frac{1}{2}(\p_{x}h)^2 + O(|\p_{x}h|^4), \nonumber \\
n_x &= &-\p_{x}h + O(|\p_{x}h|^3).
%\end{split}
%\end{equation}
\end{eqnarray}

Following an analogous procedure on the exponents, $\exp(- i k_{zp}h)$, $\exp(- i k_{z}h)$, appearing in the boundary
conditions (but in the parameter $|h|/\lambda$), after some straightforward algebra we finally obtain an expanded integral equation, in which as many
expansion terms (in powers of $ \p_{x}h$ and $|h|/\lambda$) as necessary should be retained in order to satisfy energy
conservation up to a required accuracy. However, as will be shown later, just considering up to the terms linear in $
\p_{x}h$, $|h|/\lambda$, already provides accurate energy conservation. Moreover, we will show that this even occurs
for rectangular or triangular defects which are, in principle, unfavorable cases as the shape slopes are not small
everywhere. From the physical point of view this means that small-scale spatial components of the fields do not
contribute essentially to the scattering. It is convenient to define the dimensionless wave-vector components $q=k/g$
(so that $q_{p}=k_{p}/g=\sqrt{1-\xi^2}$) and $q_z = k_{z} / g$, and the dimensionless Fourier amplitude of the relief
defect $ \eta(q)= ig^2 h(k)$. Additionally, the renormalized field amplitudes $r(q)$ are defined by
\begin{equation}\label{rq}
r(q) = g H_y(k) \, G(q)^{-1}, \q G(q)=1/(\xi + q_z),
\end{equation}
where $G(q)$ is the Green's function corresponding to the
unperturbed SPP.

Then we find that SPP scattering is governed by
\begin{equation}\label{p1}
r(q) + \int dq'\, U(q,q')G(q') r(q') = -U(q,q_{p}).
\end{equation}

In this equation $U(q,q')$ is the scattering potential which, in general, can be expressed as a series expansion in the Fourier image of the defect
profile $ \eta(q)$ as
\begin{equation}\label{p2}
 U(q,q')= U_1(q,q') + U_2(q,q')+..., \q
 U_n(q,q') \sim\eta^{n}.
\end{equation}
Up to first order in $|\p_{x} h|$, $|h|/\lambda$, the explicit expression for the potential is\cite{note1}
\begin{equation}\label{rp}
\begin{split}
U(q,q') = [q'(q-q') - q'_zG(q')^{-1}]\eta(q-q').
\end{split}
\end{equation}

We have to keep in mind that, within the ``local impedance" approximation, the SPP scattering is also governed by Eq.~\eqref{p1} but, in this case,
the scattering potential is $U_{local}(q,q') = \eta(q-q') \, (\epsilon-1)/\epsilon \approx \eta(q-q') $. This different functional form is not
irrelevant: as we will show later, the fact that $U(q_p,q_p)=0$ (property not shared by $U_{local}$) has important consequences in the scattering of
SPP by surface reliefs. In order to ascertain which is the best approximation, subsequent sections present comparisons between results obtained with
the modal expansion and both potentials $U$ and  $U_{local}$. Let us anticipate that the comparison favors the scattering potential $U(q,q')$ defined
by Eq.~\eqref{rp}.

Another point favoring $U(q,q')$ against $U_{local}(q,q')$ is to
consider the perfect conductor limit ($\epsilon \rightarrow
-\infty$), where the SIBC must transform to $(\partial / \partial n)
H_y(x,z) = 0$, evaluated at $z=h(x)$. In this limit, while Eq.
\eqref{4} transforms correctly, the use of ``local impedance" leads
[by using Eqs. (5) and (25) of Ref.~\onlinecite{MaradudinIBC}] to
the boundary condition $(\partial/ \partial z) H_y(x,z)= g^2
h(x)H_y(x,z)$, evaluated at $z=0$. The root of the problem seems to
be that the series expansion in $h(x)$ [Eq.~(2.23) in
Ref.~\onlinecite{MaradudinIBC}] diverges as
$\epsilon\rightarrow\infty$ (note, for instance, that the
second-order term is inversely proportional to the skin-depth,
$d\sim1/\sqrt{-\epsilon}$).

To complete this section, let us point out that the previously
outlined formalism can be generalized to the case of defects due
to impedance inhomogeneities. In this case, the metal surface is
flat, but the surface impedance becomes a function of the
$x$-coordinate
\begin{equation}\label{i1}
\xi(x) =\xi + \tilde{\xi}(x)= \xi + \int dk\,
\tilde{\xi}(k)\exp(ikx).
\end{equation}
After applying SIBC, which in this case reads $E_{x}(x,z=0) = \xi(x)\, H_y(x,z=0)$, we find that SPP scattering is also controlled by Eq.~\eqref{p1}.
The difference is that now the scattering potential is
\begin{equation}\label{pot_imp}
U(q,q')= \eta(q-q'),
\end{equation}
where $\eta(q) = g\tilde{\xi}(k)$ is the dimensionless Fourier
amplitude of the modulation defect.

\subsection{The transmission, reflection and out-of-plane scattering}

Once the coefficients $r(q)$ are obtained, the integrals in Eqs.~\eqref{2} define the EM field everywhere in vacuum. However, in order to find the
scattering coefficients, only the asymptotic values at large distances from the scattering center are needed. It is then convenient to prolong the
sub-integral functions into the complex plane, taking into account the presence of poles. While the renormalized Fourier image of the field, $r(q)$,
is not expected to present poles, the Green function $G(q)$ definitely does. In order to retain its causal character, an infinitesimally small
damping should be included as
 $ \xi \rightarrow i\mathrm{Im}(\xi) + 0$.
 As a result, the magnetic field at the surface $z=0$ may be written in the following form
%\begin{equation}\label{trs2}
\begin{eqnarray*}\label{trs2}
H_y(x\rightarrow\infty,0) & = &(1 + \tau) \exp(ik_{p}x ),
\\
 H_y(x\rightarrow-\infty,0) & = & \exp(ik_{p}x ) +
\rho\exp(-ik_{p}x ),
\end{eqnarray*}
 where
 \begin{equation}\label{trs3}
 \begin{split}
\tau = \frac{2\pi i\xi}{q_{p}}r(q_{p}), \q \rho = \frac{2\pi i\xi}{q_{p}}r(-q_{p}).
 \end{split}
  \end{equation}
The energy flux scattered out of the metal-vacuum interface is
computed by integrating the Poynting vector over a rectangular
parallelepiped defined by the plane $z=0$ and walls placed in the far-field parallel to the
planes $Ox$, $Oy$, and $Oz$. Then, taking
into account that the power per unit length of the incoming SPP is
$q_{p}/(4|\xi|)$, energy conservation law has the following form
\begin{equation}\label{trs4}
 \begin{split}
1-S-T-R=0,
 \end{split}
  \end{equation}
  where the reflection $R$, transmission $T$, and out-of-plane $S$ scattering coefficients are
  \begin{equation}\label{trs5}
 \begin{split}
R = |\rho|^2, \q T=|1+\tau|^2, \\ S=
\frac{4\pi|\xi|}{q_{p}}\int\limits_{|q|<1} d q\cdot q_z|G(q)r(q)|^2.
 \end{split}
  \end{equation}
The total out-of-plane scattered current can then be written in
the following form
\begin{eqnarray}\label{trs6}
S & =& \int\limits_{-\pi/2}^{\pi/2} d \theta D(\theta), \nonumber \\
D(\theta)& =& \frac{4\pi|\xi|}{q_{p}}\cos^2\theta\left|r[q(\theta)]G[q(\theta)]\right|^2 \nonumber \\
   &=&\frac{4\pi|\xi|\cos^2\theta}{q_{p}\left(|\xi|^2+
\cos^2\theta\right)}\left|r[q(\theta)]\right|^2.
\end{eqnarray}
where $D(\theta)$ is the differential reflection coefficient (DRC), which
provides the angular dependence of the radiated energy.
In this expression angle $\theta$ is defined from the
normal, while $q(\theta)=\sin\theta$ and $q_z(\theta) = \cos\theta$.

To summarize this section, the set of equations \eqref{p1},
\eqref{trs5} and \eqref{trs6} define the scattering coefficients
for an SPP impinging normally into an arbitrary set of
perturbations (with an axis of translational symmetry) in a flat
metal surface within the Rayleigh + SIBC approximations. Such
perturbations can be either variations in the surface relief
(either indentations or protrusions) or variations in the surface
impedance.

\subsection{Perturbative approach: qualitative description of the results}

In some cases, it is useful to estimate the solution of
Eq.\eqref{p1} using a perturbative approach. Taking into account the
representation of the scattering potential in \eqref{p2}, we seek
the solution of Eq.~\eqref{p1} in the form of a series expansion in
$\eta$:\footnote{Recall though that $\eta(q)$ must be a small
parameter: even for small defect depths the approximation breaks for
very stretched defects.}
\begin{equation}\label{p3}
 r(q)= r_1(q) + r_2(q) +..., \q
 r_n(q)\sim\eta^{n}.
\end{equation}
The first-order Born approximation (FOBA) gives us
\begin{equation}\label{p4}
 r_1(q) =-U_1(q,q_{p})\sim \eta(q-q_{p}).
\end{equation}
The second-order term is
\begin{equation}\label{p5}
 r_2(q) =-U_2(q,q_{p}) +\int dq'\, U_1(q,q') \, G(q') \,
 U_1(q',q_{p}).
\end{equation}
As usual, FOBA describes the scattering of the incident wave with in-plane wavevector, $q_{p}$, into a
plane wave with the wavevector $q$, the momentum difference being provided by a single interaction
with the defect. Therefore the scattering amplitude for this process is proportional
 to $\eta(q-q_{p})$. This explains the structure of the transmission coefficient.
As Eq.~\eqref{trs5} shows, $T$ has contributions from both the
 incident SPP (with unit amplitude) and from the scattering of the incident plasmon ``into
 itself'' [the term proportional to $\eta(0)$]. Analogously, in FOBA
the amplitude of the reflection is proportional
 to $\eta(-2q_{p})$ and represents the scattering of the incoming SPP into the backwardly-propagating one (having the wavevector $-q_p$).

\section{\label{sec:setup2} SPP scattering by a single defect.}

In this section we consider the dependence of scattering properties
by single 1D defects of different shapes. The defect width is $a$
and the defect depth is $\mathrm{w}$ (see Fig. 2). Notice that,
within the chosen coordinate system, indentations are characterized
by $\mathrm{w}>0$, while protrusions have $\mathrm{w}<0$. The
calculation is performed by solving numerically the integral
equation \eqref{p1}, after applying an appropriate discretization in
q-space. As the results are in good agreement with those obtained
within first-order Born approximation, analytic expressions for the
scattering coefficients are provided in several cases.

\subsection {Scattering coefficients}

\begin{figure*}[t]
\begin{center}
\epsfig{file=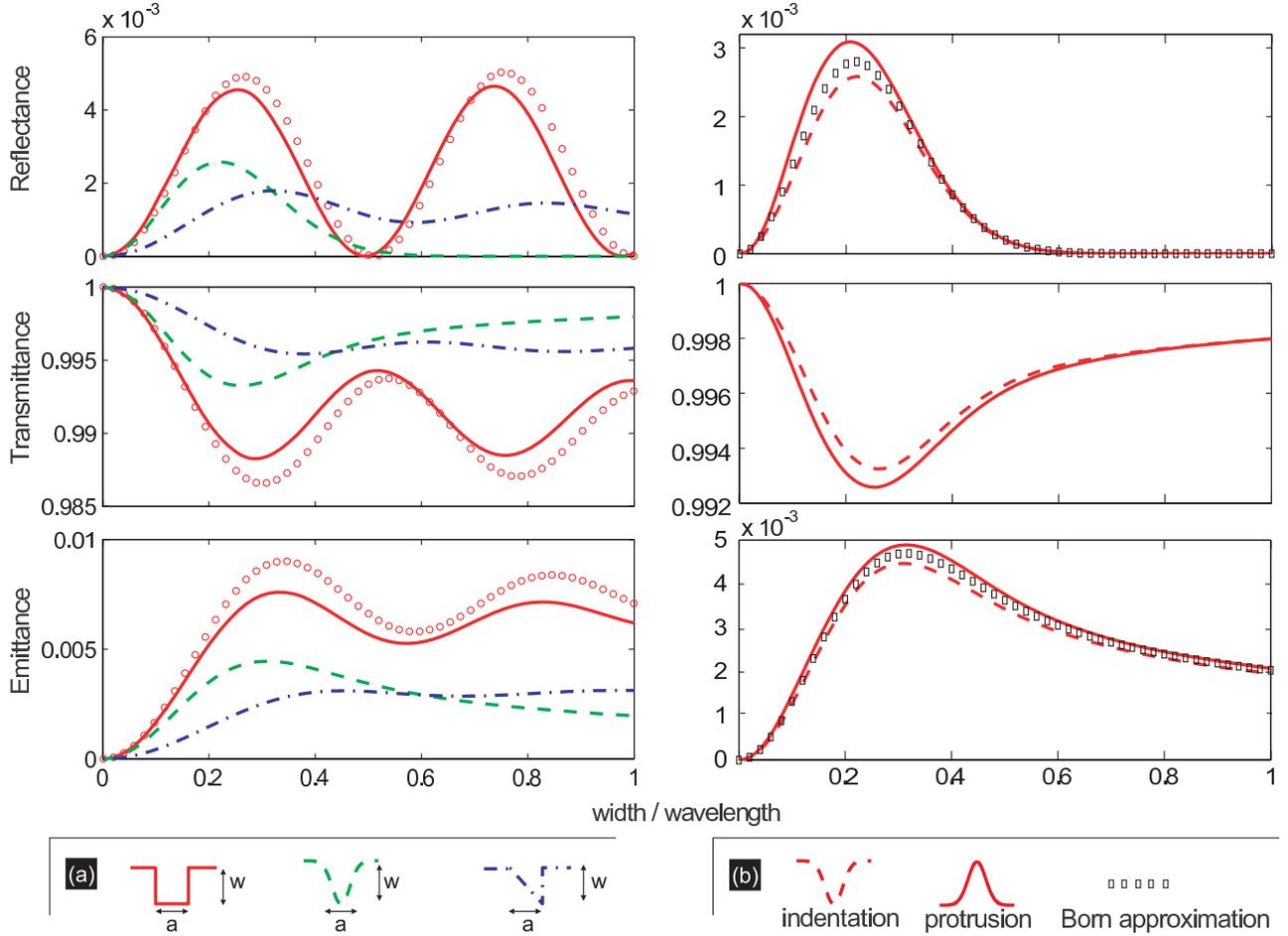,width=17cm}
\end{center}
\caption{\label{RTS} (Color online) The dependency of the
transmittance, T, reflectance, R, and emittance, S, upon the
dimensional width of the defect of different shapes in the silver
surface (at $600 nm$). The amplitude is $\mathrm{w}/\lambda = 0.02$.
In (a) solid, dashed and dash-dotted curves correspond to the
rectangular, gaussian and triangular shapes of indentation. Round
markers correspond to the calculations using mode expansion. In (b)
dashed (solid) curve corresponds to a gaussian indentation
(protrusion); by the rectangular markers the Born approximation is
presented.}
\end{figure*}

Figure~\ref{RTS} renders the results of numerical calculations for
$R$, $T$, $S$ in different instances. In (a) the comparison of these
quantities for single indentations of different shapes is presented.
In the case of a rectangular shape, $R$ behaves periodically with
respect to $a/\lambda$, while in the case of a gaussian shape $R$
possesses only one maximum. Analogously, the transmittance presents
oscillations in $a/\lambda$, in contrast with the single maximum
that appears for the case of gaussian shape. In the same figure we
present also the calculations for the triangular shape. These
results show that even shallow defects present very rich shape and
spectral dependences.

Figure ~\ref{RTS}(b) shows the scattering coefficients for both
indentations and protrusions of gaussian shape. As can be seen
from this figure, the scattering properties of indentations and
protrusions are very similar for shallow defects. Nevertheless,
for the considered parameters, protrusions present a slightly
larger cross section, resulting in larger values for both $R$ and
$S$ and smaller ones for $T$. We stress that, although the
integral equations where derived assuming that the surface was
smooth, energy conservation was fulfilled with an accuracy better
than 1\% of the minimum value of $R$, $T$ and $S$, even for sharp
defects. Additionally, we have checked that retaining second-order
terms in $\eta(q)$ in the scattering potential leaves the results
virtually unaltered. Moreover, solving the integral equation  with
the ``exact'' (not expanded) right-hand side does not produce any
significant variation in the calculated scattering coefficients.

In order to further validate the above mentioned approximations, leading to the results presented in Fig.~\ref{RTS}, additional calculations were
carried out with the modal expansion technique\cite{Luis_mirrors_05}. While also using SIBC, this technique is applicable for {\it indentations} of
any depth, going beyond the Rayleigh expansion. The comparison is presented in Fig.~\ref{RTS} (a), for the case of a rectangular indentation. As can
be seen, the agreement is very good, the difference being attributed mostly to the fact that, in the modal expansion, ideal metal boundary conditions
were used for the vertical ``walls'' of the rectangular indentation. We note in passing that using the ``local impedance" scattering potential gives
very different results: the values of $T$ and $S$ differ by more than an order of magnitude from the ones presented in Figs. \ref{RTS} and \ref{DRC} (we performed the calculations with the
potential presented in Ref.~\onlinecite{MaradudinPRB99} for the same set of parameters).

We find that FOBA provides an accurate description for the behavior of the calculated scattering coefficients. This is illustrated in Fig.~\ref{RTS}
(b), where the FOBA results for the gaussian defect are compared with the full solution of Eq.~\eqref{p1}. Notice that FOBA predicts the same
reflectance for both protrusions and indentations, since $R\sim \mathrm{w}^2$. Further analysis shows that the second order approximation already
accounts for the small differences between the scattering coefficients of indentations and protrusions found in Fig.~\ref{RTS} (b). This occurs
because, in second-order approximation $R\sim |\mathrm{w} + \mathrm{w}^2\cdot \psi(a)|^2$, where the complex function $\psi(a)$ depends upon both
shape and longitudinal size of the defect.

Besides, FOBA enables us to find analytic expressions for the
reflectance of SPPs by a single defect quite easily, in terms of
$|\eta(-2q_{p})|^2$. For instance, for a rectangular or a gaussian
defect of width $a$, we obtain
\begin{equation}
\eta(q)^{Rect}=i\Delta\frac{\sin(q \tilde{a})}{\pi q},\,
\eta(q)^{Gauss}=\frac{i\tilde{a}\Delta}{2\sqrt{\pi
}}e^{-q^2\tilde{a}^2/4},
\end{equation}
where $\tilde{a} = a\pi/\lambda$ and $ \Delta = g\mathrm{w}$ for
relief defects (for impedance defects $\Delta$ corresponds to the
maximum value of $|\tilde{\xi}|$). Therefore, the reflectance of a
single relief defect can be expressed as
\begin{eqnarray}\label{R_FOBA}
R^{Gauss} & = & 4\pi|\xi|^2 q_p^2
\Delta^2\tilde{a}^2e^{-2q_p^2\tilde{a}^2},
\nonumber\\
 & & \\
R^{Rect} & = & 16|\xi|^2 q_{p}^2\Delta^2\tilde{a}^2
\mathrm{sinc}^2\left(2 q_{p} \tilde{a}\right). \nonumber
\end{eqnarray}

Thus, for rectangular defects $R$  behaves periodically as a function of $a/\lambda$, possessing minima at $a/\lambda =
n/2q_{p}$, $n=1,2,...$; while the reflectance for defects of gaussian shape presents only one maximum at $a/\lambda =
1/\sqrt{2}\pi q_{p}$. This is in an excellent accordance with the strict numerical solution of Eq.~\eqref{p1}, see
Fig.~\ref{RTS} (a), (b). However, as the transmittance in FOBA does not depend upon the width of the defect, higher
order terms in the Born series are required in order to reproduce this dependence appropriately.

\subsection{Out-of-plane radiation due to a single defect.}

\begin{figure}[h!]
\begin{center}
\epsfig{file=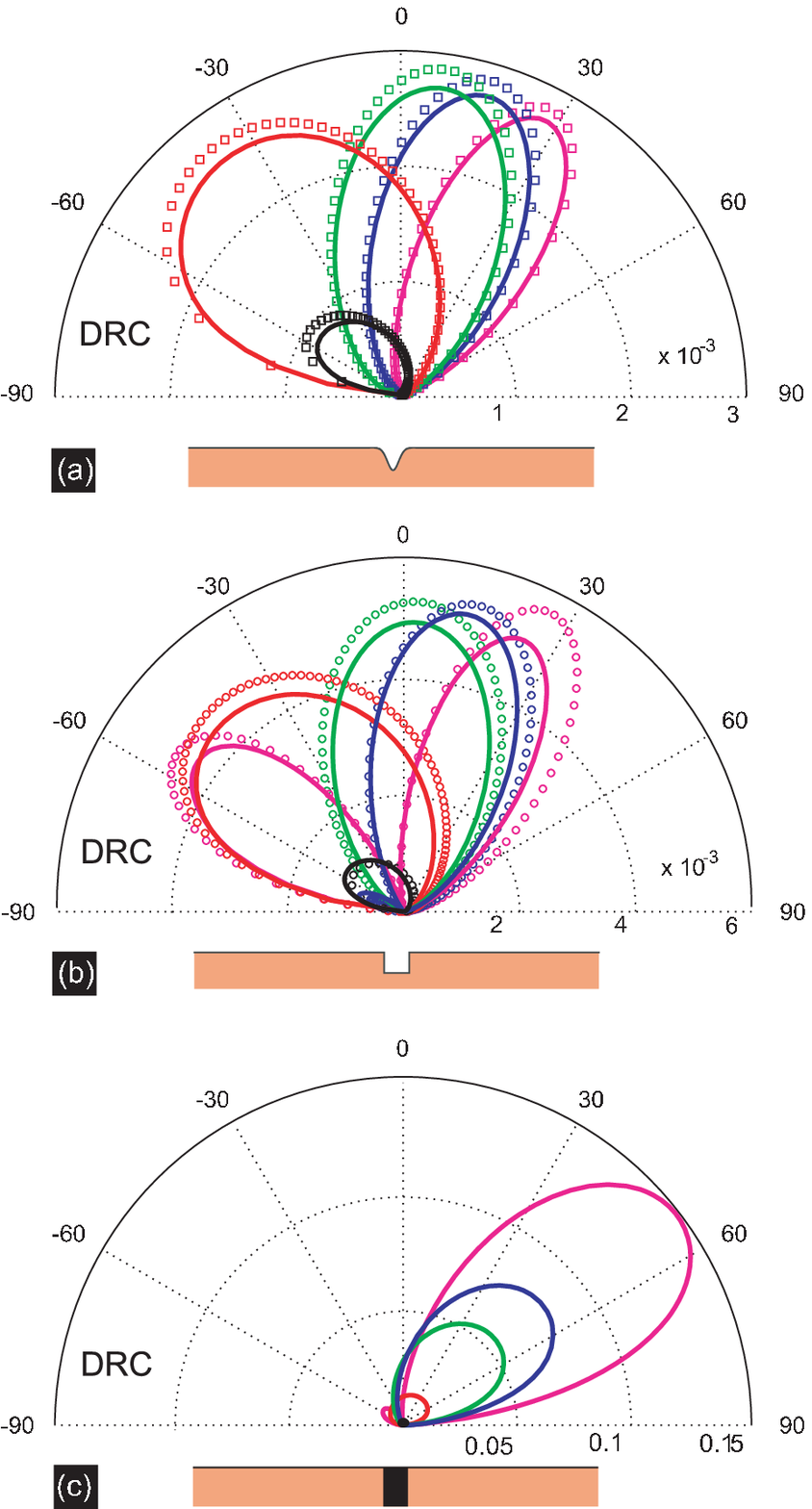,width=\columnwidth}
\end{center}
\caption{\label{DRC} (Color online) DRC, $D(\theta)$, for SPP scattering
along a silver surface ($\lambda=600 nm$), for different defect
types and widths. (a) gaussian indentation, (b) rectangular
indentation and (c) impedance step (the metal is considered ideal
inside the defect). Dash-double-dotted (black), dashed (red),
dotted (green), dash-dotted (blue), solid (magenta) correspond to
defect widths $a/\lambda= 0.1$, $0.25$, $0.5$, $0.6$, $0.8$,
respectively. The depth of the indentation is $\mathrm{w}/\lambda
=0.02$. Squares in (a) correspond to calculations within the first
order Born approximation. Round markers in (b) correspond to the
calculations using modal expansion}
\end{figure}

\begin{figure*}[]
\begin{center}
\epsfig{file=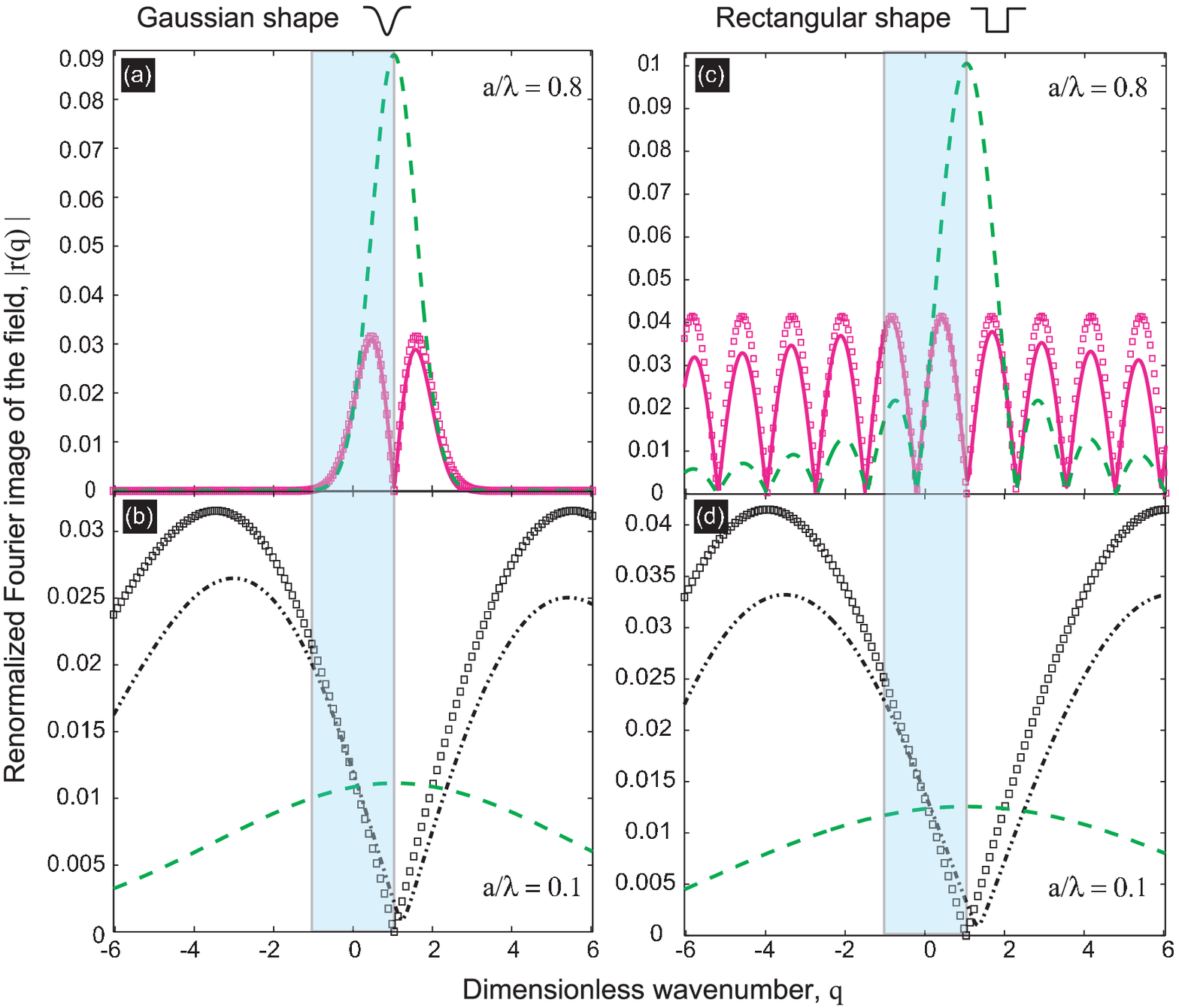,width=16cm}
\end{center}
\caption{\label{FFT}(Color online) Dependence of the modulus of the
renormalized Fourier image of the scattered field, $|r(q)|$, upon
dimensionless wave-vector, $q$. Left panels show the calculation
for gaussian indentations [$a/\lambda = 0.8$ in panel (a),
$a/\lambda = 0.1$ in panel (b)], while right panels correspond to
rectangular indentations [$a/\lambda = 0.8$ in panel (c),
$a/\lambda = 0.1$ in panel (d)]. The depth of the indentation is
$\mathrm{w}/\lambda =0.02$. In each panel the squares represent
the $|r(q)|$ calculated within first order Born approximation,
whereas dash-double-dotted (black) and solid (magenta) lines
stands for the strict numerical solution; the green dashed curve
renders the modulus of the $q_p$-shifted Fourier transformation of
the defect, $|\eta(q-q_{p})|$. Notice the dips of all curves at
$q=q_{p}$ due to the presence in the potential of the ``cutting''
function $|q-q_{p}|$. The blue shaded areas denote the regions in
$q$ corresponding to out-of-plane radiative modes. }
\end{figure*}

In this section we analyze the angular distribution of the the energy radiated out of the plane after scattering. This behavior is represented in
Fig.~\ref{DRC}, which renders the radiation diagrams for SPP scattering by gaussian indentations [panel (a)] and rectangular indentations [panel
(b)], for different defect widths. The surface impedance is that of silver at $\lambda=600$nm ($\xi=-0.277i$). The case of an impedance defect (with
zero impedance, as for perfect conductors) is also shown in Fig.~\ref{DRC}~(c). For comparison, the calculations for rectangular indentations were
also performed with the modal expansion method [circles Fig~\ref{DRC}(b)]. Again, the agreement between the two methods is quite remarkable.
Fig.~\ref{DRC} shows that the radiation diagrams present a non-trivial dependence on defect shape. For impedance defects, the angle at which maximum
out-of-plane radiation occurs always points along the direction of propagation of the incident SPP. However, for narrow gaussian relief defects
($a<\lambda/2$), the maximum in DRC occurs at negative angles. In this case, as $a$ increases, the angle of maximal radiation shifts from negative
angles to positive ones. This behavior of gaussian defects has also been reported for surface plasmons in a {\it thin metal film}, excited in a
Kretschmman configuration\cite{MexicansPRB06}. For rectangular defects, the DRC behavior is more complex. At small $a/\lambda$ the DRC presents one
emission lobe at a negative angle; as $a$ increases, the emission lobe moves to the normal direction and after a transition point (at approximately
$a/\lambda$ = 1/2) the main lobe moves to positive angles, while a second emission lobe appears at a negative angle. Finally, for $a/\lambda \approx
1$ the amplitudes of the two lobes are comparable, see Fig.~\ref{DRC}.

Within FOBA, the DRC can be analytically computed. From
Eq.~\eqref{p4} and Eq.~\eqref{trs6} we obtain
\begin{equation}\label{d2}
 \begin{split}
D_1(\theta)=\frac{4\pi q_{p}|\xi|(\sin\theta-q_{p})^2\cos^2\theta}{
|\xi|^2+\cos^2\theta }\left| \eta(\sin\theta-q_{p})\right|^2.
 \end{split}
\end{equation}
As illustrated in Fig.~\ref{DRC} (a) for the case of gaussian indentations, FOBA results (square symbols) provide and excellent approximation to the
full solution. This allows us to connect the DRC with the potentials corresponding to the SPP scattering within such approximation.

For the case of impedance defects, Eq.~\eqref{pot_imp} and
Eq.~\eqref{p1} imply that the renormalized spectra of the field
$r(q)$ is proportional to $\eta(q-q_p)$, which is the Fourier image
of the defect shifted by the SPP wave-vector\footnote{Evidently, the
true spectra of the field, $H_y(k)$, is also strongly influenced by
the SPP Green's function, see Eq.~\eqref{rq} }. Conversely, for
relief defects, $r(q)$ is proportional not only to $\eta(q-q_{p})$
but also to $q-q_{p}$. This is illustrated in Fig.~\ref{FFT}, for
gaussian (left panels) and rectangular (right panels) at
$a/\lambda=0.8$ (top panels) and $a/\lambda=0.1$ (bottom panels). In
all panels the modulus of the shifted Fourier image of the defect
$|\eta(q-q_p)|$ is represented by the green dashed lines.
Fig.~\ref{FFT} also shows the corresponding $|r(q)|$ for relief
defects, obtained by solving the integral equation  \eqref{p1}
either strictly or within FOBA approximation.

As expected, the narrower the defect, the more extended
$|\eta(q-q_p)|$ (and therefore $|r(q)|$) in $q$-space. It is also
clear that $|\eta(q-q_p)|$ is a smooth function for gaussian
defects, while it presents oscillations for rectangular ones.
These two properties are extremely useful in order to understand
the behavior of DRC. Notice that $D(\theta)$ essentially follows
the dependence of $|r(q)|^2 $ (see Eq.~\ref{trs6} within the range
$-1<q<1$ i.e., the region corresponding to radiative modes,
represented by the blue shaded areas in Fig.~\ref{FFT}). For
impedance defects $|\eta(q-q_p)|$ presents its maximum at $q>0$,
so the maximum out of plane emission always occurs in the forward
direction, exactly in the same way as in the model of the
\emph{local} surface impedance
approximation\cite{MaradudinIBC,SanchesAppl.Lett.98,MaradudinPRB99}.
However, for relief defects, forward direction emission is further
inhibited by the presence of the ``cutting function" $q-q_{p}$ in
the expression for $r(q)$. For narrow defects this results in
$|r(q)|$'s having more weight on the $q<0$ region, so the emission
is directed backwards. For wide gaussian defects ($a> 0.5\lambda$)
the width in q-space of $|\eta(q)|$ is smaller than $q_p$ (recall
that $q_p \approx 1$) and the emission occurs in the forward
direction. For rectangular defects, as the defect widens, more and
more oscillations in $|\eta(q-q_p)|$ enters into the $-1<q<1$
region, leading to different emission lobes, as shown in
Fig.~\ref{DRC}(b).

\begin{figure}[!h]
\begin{center}
\epsfig{file=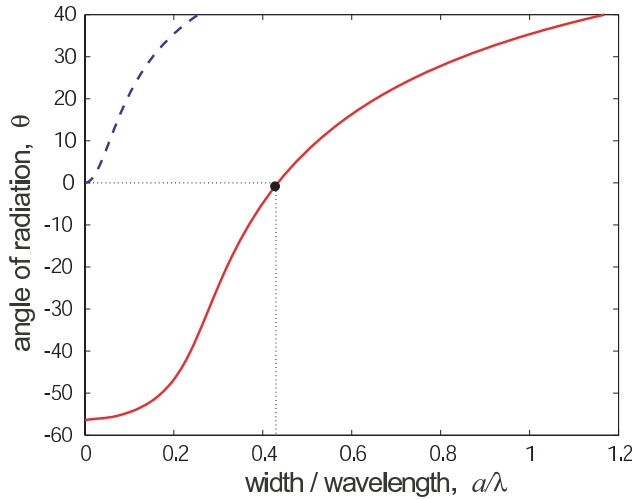,width=\columnwidth}
\end{center}
\caption{\label{curve_max}(Color online) Dependence of angle of
maximum out of plane radiation as a function of $a/\lambda$ for
gaussian defects. The case of relief defects is represented in the
red continuous curve, the blue broken curve corresponds to impedance
defects.}
\end{figure}

The angles of maximum out-of-plane emission can be obtained in
FOBA by solving $\p_{\theta}D_1(\theta)=0$. While in the case of a
rectangular shape this equation is transcendent with respect to
variables $\theta$ and $a/\lambda$, it may be solved explicitly
for gaussian shapes. Thus, we find that the maxima condition for
gaussian reliefs is
\begin{equation}\label{d3}
\frac{a}{\lambda}(\theta) =
\sqrt{\frac{2}{\pi^2(\sin\theta-q_{p})^2}+\varphi(\theta)},
\end{equation}
where \begin{equation}
 \q \varphi(\theta) =
\frac{2|\xi|^2\tan\theta}{\pi^2\left(|\xi|^2+\cos^2\theta\right)(q_{p}-\sin\theta)\cos\theta},
\end{equation}
while in the case of an impedance defect this condition is much
simpler, $a/\lambda\,(\theta)=\sqrt{\varphi(\theta)}$. These
dependencies are presented in Fig.~\ref{curve_max}, reflecting
that while impedance defects emit preferentially in the forward
direction, emission maxima occurs in the backward direction up to
$a/\lambda \approx 0.42$ for the relief case .

\section{\label{concl}Conclusions}

We have developed an approximate method (based on Rayleigh
expansion and surface impedance boundary conditions) for the study
of SPP scattering at arbitrary 1D inhomogeneities, and applied it
to the case of single defects.

We have compared the numerical solution of the integral equation in
$k$-space with the calculations made by using mode expansion
technique. The excellent agreement between them (together with the fulfillment of
the energy conservation law) indicates that the theoretical
formulation is correct and much more accurate than previous approximations
based on {\it local} surface impedances.

The case of a single scattering center is analyzed in detail. We
have compared the scattering by impedance and relief inhomogeneities of different
shapes, considering both protrusions and indentations. We have shown that the
transmission, reflection and out-of-plane scattering, are
defined essentially by the spectral properties of the inhomogeneity.
We have shown that, out-of-plane radiation after scattering
by impedance inhomogeneities is always directed in the forward direction
(with respect to the SPP propagation). On the contrary,
in the case of relief defects, the radiated energy may be directed
into backward or forward directions (or both, in the case of rectangular defects),
depending on the defect width and shape.

Further theoretical work will be aimed at the scattering of SPP by the multiple scatterers,  and, also, at the generalization of this approach to
$2D$ inhomogeneities.

\section{\label{ackn}Acknowledgments}

We are grateful to Prof. A.V. Kats and Dr. J. A. S\'{a}nchez-Gil for helpful discussions and criticism. The authors acknowledge financial support
from the European Network of Excellence Plasmo-Nano-Devices (FP6-2002-IST-1-507879) and the STREP ``Surface Plasmon Photonics''
(FP6-NMP4-CT2003-505699).


\begin{thebibliography}{99}
\bibitem{note1} Everywhere we take into account that, due to absence of dissipation, $G(q_{p})^{-1}=0$.
\end{thebibliography}
\end{document}